\documentclass{article}

\newtheorem{example}{Example}
\usepackage{color}
\usepackage{graphicx}
\usepackage{subfig}
\usepackage{url}
\usepackage{relsize}

\usepackage{amsmath}

\usepackage{MnSymbol}%

\usepackage{float}
\usepackage{wrapfig}
\usepackage{url}
\usepackage{xspace}
\usepackage{anysize}
\usepackage{todonotes}
\usepackage[plain]{algorithm}
\usepackage{multirow}

\usepackage{bm}

\usepackage{algorithm}
\usepackage{algorithmicx}
\usepackage{algpseudocode}

\algtext*{EndIf}%

\usepackage{listings}
\usepackage{bold-extra}

\definecolor{dkgreen}{rgb}{0.5,0.5,0.5}
\definecolor{gray}{rgb}{0.5,0.5,0.5}
\definecolor{mauve}{rgb}{0.1,0.5,0.1}
\definecolor{lightgray}{rgb}{0.95,0.95,0.98}
\definecolor{keyw2}{rgb}{0.75,0.1,0.1}
\definecolor{keyw}{rgb}{0,0,0.4}
\lstdefinelanguage{scala}{
  morekeywords={abstract,case,catch,class,def,%
    do,else,extends,false,final,finally,%
    for,if,implicit,import,match,mixin,%
    new,null,object,override,package,%
    private,protected,requires,return,sealed,%
    super,this,throw,trait,true,try,%
    type,val,var,while,with,yield},
  otherkeywords={=>,<-,<\%,<:,>:,\#,@},
  sensitive=true,
  morecomment=[l]{//},
  morecomment=[n]{/*}{*/},
  morestring=[b]",
  morestring=[b]',
  morestring=[b]"""
}

\papersize{26.7cm}{20.7cm}
\usepackage{tikz}
\definecolor{darkgreen}{rgb}{0,0.66,0} 
\usepackage{pgfplots}
\usetikzlibrary{arrows}

\usepackage{ifthen}
\usepackage{calc}

\date{June 20th, 2014}

\newcommand{\framework}{{\rmfamily\scshape FooPar}\xspace}

\begin{document}

\title{Group Communication Patterns for \\High Performance Computing in Scala}

\author{Felix P. Hargreaves\\ Daniel Merkle \\ Peter Schneider-Kamp\\[0.3cm]
           Department of Mathematics and Computer Science\\ University of
Southern Denmark\\
           \{\texttt{hargreaves,daniel,petersk\}@imada.sdu.dk}}

\maketitle

\begin{abstract}

We developed a {\rmfamily\scshape F}unctional {\rmfamily\scshape
o}bject-{\rmfamily\scshape o}riented {\rmfamily\scshape Par}allel framework
(\framework) for high-level high-per\-formance computing in Scala.
Central to this framework are Distributed Memory Parallel Data
structures (DPDs), i.e., collections of data distributed in a shared nothing
system together with parallel operations on these data.

In this paper, we first present \framework's architecture and the idea of DPDs
and group communications. Then, we show how DPDs can be implemented elegantly
and efficiently in Scala based on the \textit{Traversable/Builder} pattern, unifying Functional and Object-Oriented Programming.

We prove the correctness and safety of one communication algorithm and show how specification testing (via ScalaCheck) can be used 
to bridge the gap between proof and implementation. Furthermore, we show that the group communication operations of \framework outperform
those of the MPJ Express open source MPI-bindings for Java, both asymptotically and empirically.

\framework has already been shown to be capable of achieving close-to-optimal performance for dense matrix-matrix multiplication via JNI. 
In this article, we present results on a parallel implementation of the Floyd-Warshall algorithm in
\framework, achieving more than 94\% efficiency compared to the serial version on
a cluster using 100 cores for matrices of dimension $38000 \times 38000$.
\end{abstract}

\section{Introduction}

Building scalable systems in the presence of several thousands, even millions,
of computational cores asks for new programming paradigms in order to
meet obvious goals such as performance exploitation, correctness, portability,
fault tolerance, and usability. It is well foreseeable that the
programming model underlying the Message Passing Interface (MPI),
the current de-facto standard for programming distributed
memory systems, will not be able to address the challenges for the
necessary next steps in High Performance Computing (HPC). 
 Reasons for
this include that MPI ignores memory hierarchies and that reaching a
high productivity (i.e., performance, expressivity, portability, and
robustness)
is challenging based on the error-prone low-level programming
model \cite{ctwatch06}.

It is no surprise that new programming languages
for simplifying the programming on peta- and exascale parallel systems
are currently developed and studied. IBM's X10 \cite{x10}, Cray's Chapel~\cite{chapel}, and
Sun's Fortress \cite{Alle07a} are three prominent examples, that are
based on the partitioned global address space (PGAS) programming
model. While PGAS
aims at combining the advantages of using an SPMD programming model on
distributed memory systems with the advantages of using referencing
semantics of shared memory systems, this approach also comes with some drawbacks.
To reach high performance users still need to
specify and reason about the placement of data and tasks, e.g., in
Chapel a {\em local} type is used for that and in X10 user-defined
{\em places} encapsulate binding of activities and globally
addressable data. Also, proving correctness or analyzing
scalability is usually not considered a prominent design goal.

A promising approach for HPC comes from Delite \cite{rompf11}, a framework and runtime for the definition of 
high performing DSLs, encompassing techniques such as loop-fusion, common subexpression elimination, term rewriting, modular code emitters (support for SIMD instructions), automatic 
parallelization of expressions based on analysis of side-effects, and much more. The syntactical part of the DSLs then relies on implementations using \textit{lightweight modular staging} \cite{rompf13}, an
advanced design pattern utilizing a very minimal compiler plugin for Scala. This approach solves many problems, but is ultimately not analyzable from the end 
user perspective, i.e., it can become difficult to gauge the running time of programs written in these DSLs. Additionally, while this approach is by far more productive than building 
external DSLs from scratch, it requires a big skill set from the DSL author.

While the goal of our work is also a high productivity approach for HPC, our
contribution is orthogonal to the aforementioned approaches. In this
paper we describe \framework, a parallel framework which is based on
the functional object-oriented programming language Scala \cite{scalref}. Functional
languages have at most played a niche role in HPC, often both due to
acceptance problems\cite{wadler98} and due to an inability to perform on-par with
hand-optimized C code close to the theoretical limits of the
hardware. \framework heavily employs distributed memory data
structures especially suited for HPC, enabling it to achieve
excellent performance. It offers a
balance between conciseness and performance, achieving scalability and
efficiency close to that of hand-optimized C code from a handful of lines of Scala 
code combined with native interface computational kernels (cf.\ Figure \ref{fig:efficiency}, data from \cite{ppam13}).
However, \framework does not only allow for high performance from an
empirical point of view, but also allows for a theoretical scalability
analysis. In addition, our framework allows for correctness proofs
of parallel code as well as specification testing \cite{odersky10}.

The paper is structured in the following
way. Section~\ref{sec:framework} describes the rationale behind the
framework and compares it with further related work. Then,
Section~\ref{sec:dpd} introduces the concept of Distributed memory
Parallel Data structures (DPDs) and their associated group
communication operations including information about parallel runtimes useful for analysing
code written in \framework. The implementation of these data structures
based on Scala's functional object-oriented features is presented in
Section~\ref{sec:implementation}. In Section~\ref{sec:verifytest} we demonstrate how correctness can be
proven for the algorithms used in \framework and how specification testing
can be used to ensure reliability of their implementation.
In Section~\ref{sec:empirical} we 
provide empirical evidence for the efficiency and scalability of our
framework. Finally, we conclude briefly in
Section~\ref{sec:conclusion}.

\begin{figure}[t]
\centering
\small
\begin{tabular}{|l|l||c|c|}\hline
 \textbf{\#cores} & \textbf{$\bm{k}$} &\textbf{optimized C}  &
\textbf{\framework} \\\hline
 216 & $30240$& 26.72 &27.07 \\\hline
 343 & $30240$& 17.27 & 17.58   \\\hline
 512 & $30240$& 11.50 & 12.51   \\\hline
 512 & $40000$& 25.21 & 26.40 \\\hline
\end{tabular}
\caption{\it\small Runtimes in seconds
for $k \times k$ matrix-matrix multiplication
using \textbf{\framework} with Open MPI Java bindings and \textbf{optimized C} with native Open MPI
on the \textit{Carver cluster} at NERSC \cite{urltocarver}. 
For $k = 40000$, \framework reaches 4.84 TFLOP/s, i.e., 88.8\%
efficiency w.r.t. the theoretical peak performance.} 
\label{fig:efficiency}
\end{figure}

\section{The \framework Framework}
\label{sec:framework}

The main goal of \framework is to avoid common challenges of distributed memory parallel programming and High Performance Computing (HPC) through the use of
high-level abstractions while maintaining analyzability:\\[-3ex]
\begin{itemize}
\item Using functional programming concepts, the \emph{Single Program Multiple Data} (SPMD) \cite{spmd} concept can be combined with \emph{Single Instruction Multiple Data (SIMD)} at a data structure level. This allows algorithms to be
formulated in virtually the same way as their serial versions (see Example~\ref{ex:ppi} below).
\item We abstract away peer-to-peer message passing by introducing a set of group communication operations appropriate for HPC use. In this way, we can avoid deadlocks, starvation, race conditions, and other common concurrency issues.
\item We avoid the many pitfalls of manual memory management through the use of a managed programming language running on top of the Java virtual machine. In addition, this provides platform independence.
\end{itemize}

\subsection{Design and Related Work}
While \framework complements the \emph{parallel collections} \cite{odersky11} introduced in Scala
2.8, it is not meant as an extension. This is due to multiple reasons. First,
the parallel collections use workload-splitting strategies leading to communication bottle-necks in distributed memory settings.
Second, they employ an implicit master-slave paradigm unsuitable for massively distributed HPC. Third,
the SPMD paradigm requires launching multiple copies of the process as opposed to branching internally into threads.

\framework differs from other functional programming frameworks for parallel computations in some key aspects. While frameworks like Eden \cite{eden05}, Spark \cite{spark10}, and Scala's own parallel collections \cite{odersky11} try to maximize the level of abstraction, 
this is mostly done through strategies for data-partitioning and distribution which in turn introduce network and computation bottlenecks. Furthermore, these tools lend themselves poorly to parallel runtime analysis hindering 
asymptotic guarantees that might otherwise be achieved. To unaware users, ``automagic'' parallel programming can easily lead to decreased performance due to added overhead and small workloads. With this in mind, 
\framework aims at the sweetspot between high performance computing and highly abstract, maintainable and analyzable programming. This is achieved
by focusing on user-defined workload distribution and deemphasizing fault tolerance. In this way, the performance pitfalls of both \emph{dynamic workload allocation} and the \emph{master-slave} paradigm can be avoided
and \framework can provide HPC parallelism with the conciseness, efficiency and generality expected from mature Scala libraries, while nicely complementing the existing parallel collections of
Scala's standard API for shared memory use.

While Scala's parallel collections are limited to shared memory systems, \framework works both in shared nothing as well as shared memory architectures. Taking some inspiration from MPI~\cite{gabriel04}, \framework implements 
most of the essential operations found in MPI in a more convenient and abstract level as well as expanding upon them. As an example, \framework supports reductions with arbitrary types and variable sizes, e.g. reduction 
by list or string concatenation is entirely possible and convenient in \framework (however inherently unscalable). As an addition, performance impact from the use of concatenation or other size-increasing operations is directly visible through the 
provided asymptotic runtime analysis for operations on the Distributed Memory Parallel Data Structures (cf.\ Section~\ref{sec:dpd}). 

\framework shares goals with the partitioned global address space (PGAS) programming model in the sense that the reference semantics of shared memory systems is combined with the SPMD style of programming. Prominent examples of PGAS are Unified Parallel C, or Co-Array Fortran among others \cite{pgas05}. Focusing on performance and programmability for next-generation architectures, novel languages like X10 and Chapel provide richer execution frameworks and also allow asynchronous creation of tasks \cite{apgas07}. All these languages either resemble and extend existing languages or are designed from scratch; their features are usually accessed via syntactic sugar.
\framework, in contrast, is more oriented towards abstraction by employing distributed data structures and combining this with the mathematical abstraction inherently integrated in functional languages like Scala. This approach 
is somewhat similar to that of STAPL \cite{rauchwerger10}, however, the combination with functional programming has the potential to be more productive and produce more analyzable code.

Finally, comparing to frameworks based on Multiple Program Multiple Data (MPMD), the SPMD paradigm used in \framework emphasizes rank-data mapping, 
where ranks are the IDs of the processing elements in an execution. Section \ref{sec:dpd} shows how
rank-data mappings play a major role in \framework and how it can abstract over serial and parallel programming.

\subsection{Communication Groups}

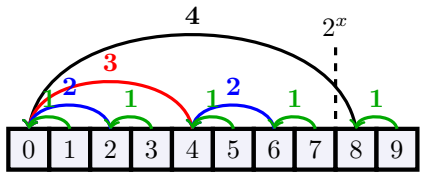
\begin{figure}[h]
 \centering
\begin{tikzpicture}

\def \u {0.9}
\def \bs {\u*0.6}
\def \n {9}
    \foreach \y in { 0,...,\n} 
    {
    \draw [fill=lightgray,ultra thick] (\y*\bs,0) rectangle(\bs*\y+\bs,\bs);
    \node at (\y*\bs + 0.5*\bs,0.5*\bs) {\y};
    }
    
\node at (8*\bs,3.5*\bs) {$2^x$};
\draw[very thick,-,dashed] (8*\bs,3*\bs) to (8*\bs,1*\bs);

\tikzstyle{every node}=[font=\bf]

\draw[very thick,<-,black] (0.5*\bs,\bs) to  [bend left=80]
node[midway,above,black]
{\textcolor{black}{4}} (8*\bs+\bs*0.5,\bs);
    
\draw[very thick,<-,red] (0.5*\bs,\bs) to  [bend left=80]
node[midway,above,black]
{\textcolor{red}{3}} (4*\bs+\bs*0.5,\bs);

\draw[very thick,<-,blue] (0.5*\bs,\bs) to  [bend left=80]
node[midway,above,black]
{\textcolor{blue}{2}} (2*\bs+\bs*0.5,\bs);
\draw[very thick,<-,blue] (4*\bs+0.5*\bs,\bs) to  [bend left=80]
node[midway,above,black]
{\textcolor{blue}{2}} (6*\bs+\bs*0.5,\bs) ;

\draw[very thick,<-,darkgreen] (0.5*\bs,\bs) to  [bend left=80]
node[midway,above,black]
{{\textcolor{darkgreen}{1}}} (1*\bs+\bs*0.5,\bs);
\draw[very thick,<-,darkgreen] (4*\bs+0.5*\bs,\bs) to  [bend left=80]
node[midway,above,black]
{{\textcolor{darkgreen}{1}}} (5*\bs+\bs*0.5,\bs) ;
\draw[very thick,<-,darkgreen] (6*\bs+0.5*\bs,\bs) to  [bend left=80]
node[midway,above,black]
{{\textcolor{darkgreen}{1}}} (7*\bs+\bs*0.5,\bs) ;
\draw[very thick,<-,darkgreen] (2*\bs+0.5*\bs,\bs) to  [bend left=80]
node[midway,above,black]
{{\textcolor{darkgreen}{1}}} (3*\bs+\bs*0.5,\bs) ;

\draw[very thick,<-,darkgreen] (8*\bs+0.5*\bs,\bs) to  [bend left=80]
node[midway,above,black]
{{\textcolor{darkgreen}{1}}} (9*\bs+\bs*0.5,\bs) ;

\end{tikzpicture}
\caption{\it 10-node reduction by inverse recursive doubling; numbers on the edges indicate the time step of the reduction.}
\label{fig:invdoubling}
\end{figure}

Instead of explicit message passing, we use the notion of \textit{groups},
a collection of \textit{processes} with a set ofoup
communication operations. We view the instantiation of a DPD of type $T$ as a projection from the set of all processing elements, $P$, defined as $P \xrightarrow{\text{subgroup}} P \xrightarrow{\text{data}} T$.
Subgroup mappings allow for communication algorithms to work independently on a multitude of topologies and sets of processing elements without cluttering the implementations 
with special cases.
Figure \ref{fig:invdoubling} shows a parallel reduction algorithm following the
pattern of the \textit{recursive doubling algorithm} \cite{kumar09}. Here, associativity of the reduction operator is assumed.
This group communication pattern can be directly mapped to the reduction operation
on a DPD as shown in Section~\ref{sec:implementation}. 
Using \framework, complete parallel computations can be described concisely as
a chain of such operations indirectly invoked through operations
on DPDs.\\[-2.5ex]

\begin{figure}[t]
\begin{center}
\hspace{-1ex}
 \includegraphics[scale=0.63]{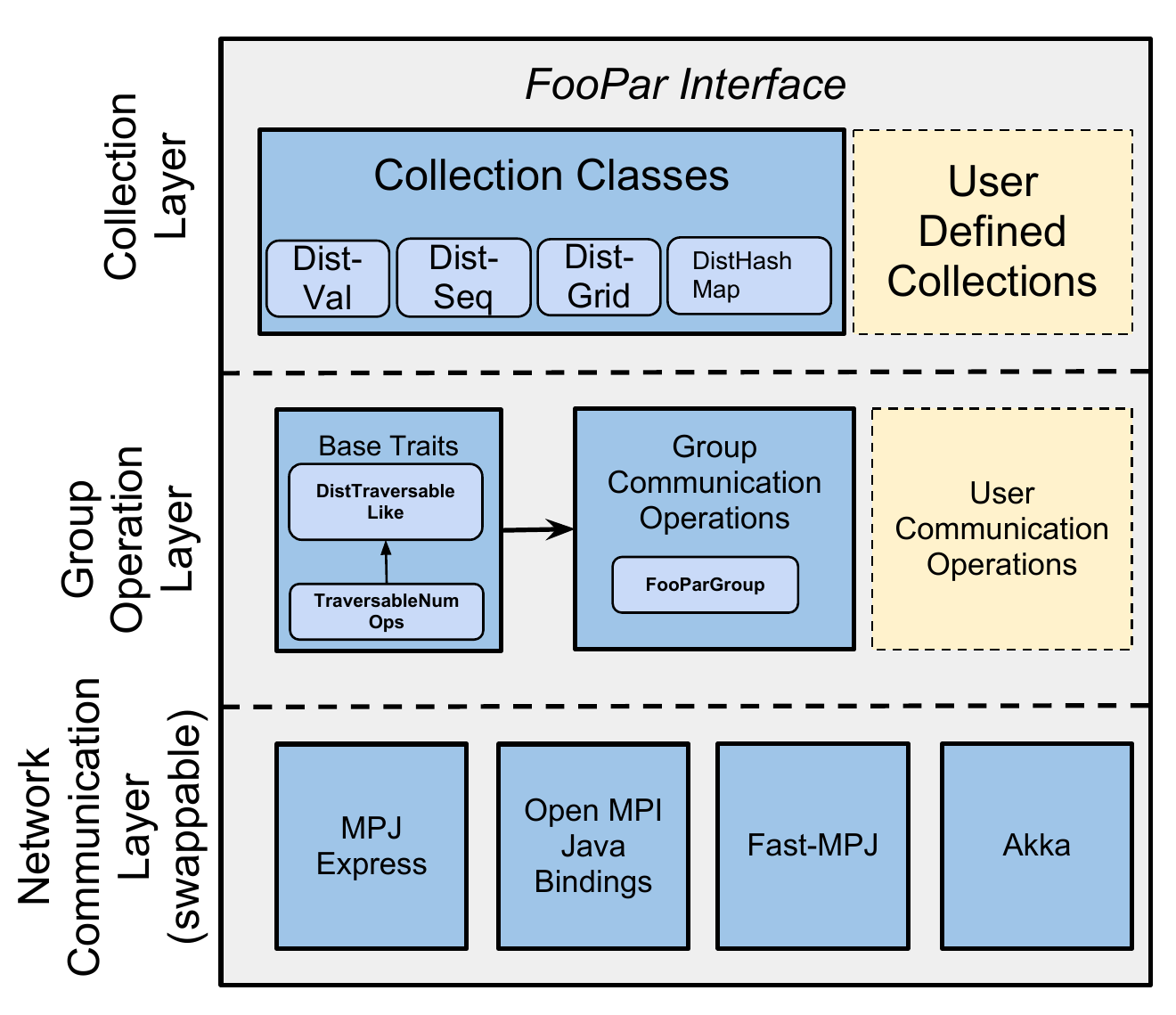}
 \caption{\it \framework's layered architecture}
 \label{fig:layered}
\end{center}
\end{figure}

\subsection{Architecture}

Figure \ref{fig:layered} shows the three layers of \framework's architecture: on
top the
DPDs, in the middle the group communication operations, and on the bottom
the backends
that abstract from network and communication specifics.

At the beginning of an execution, to facilitate this abstraction, a \textit{provider} of communication groups and processing elements is chosen
and instantiated. In this way, both industry-proven libraries,
such as \textit{MPI} \cite{gabriel04,shafi09,taboada_jos10},
and an abstract implementation provided by
\framework with arbitrary network backends (e.g. \textit{Akka}~\cite{rb12}) can be used. 
The communication patterns work across multiple communication backends due to
the \textit{process} interface, which abstracts the
message-passing functionality.

Finally, the user is presented with an SPMD programming model using DPDs
with SIMD-like \cite{spmd} operations. This combination provides a powerful
abstraction of parallelism while maintaining enough control for writing efficient and
analyzable programs.

\begin{example}
\label{ex:ppi}
Consider the following \emph{embarrassingly parallel} algorithm for approximating
the transcendental constant $\pi$ with arbitrary precision, here expressed as a
Scala function approximating the integral $\int_0^1\frac{4}{1+x^2} dx$.
{\normalfont
\begin{lstlisting}[language=scala]
def pi(n: Int) = {
  val f = (x: Double) => 4d / (1d + x * x)
  val ff = (x: Int) => (x - 0.5d) / n
  (1 to n).map(f compose ff).sum / n
}
\end{lstlisting}
}%
\noindent
Line 1 defines a function \verb!pi! taking an integer (the number of "samples" from the
integral), Line 2 defines the function \verb!f! to integrate, and Line 3 defines the function \verb!ff!
generating the sample positions. Line 4 uses a Scala built-in to create the list \verb![1,2,...,n]!,
maps the composition of \verb!f! and \verb!ff! to each element of this list, and reduces the
resulting list to its average, which is returned by \verb!pi! as its value.

In order to obtain a parallel version, we simply replace Scala's built-in lists in Line 4 with \framework's DPD for sequences, {\normalfont DistSeq},
and use the DPD methods {\normalfont mapD} and {\normalfont avgD} which correspond to {\normalfont map} and {\normalfont sum / n}, respectively:

{\normalfont
\begin{lstlisting}[language=scala]
def pi(n: Int) = {
  val f = (x: Double) => 4d / (1d + x * x)
  val ff = (x: Int) => (x - 0.5d) / n
  DistSeq.ranged(1 to n).mapD(f compose ff).avgD
}
\end{lstlisting}
}
\noindent
The sequential runtime $T_S$ is obviously in $\Theta(n)$.
For the parallel runtime $T_P$, assuming $p$ processing
elements, we
have $T_P \in \Theta(n/p + T_C)$ where
$T_C$ is the communication time.
A naive implementation of the sum
would require linear communication overhead (i.e., $T_C \in \Theta(p \cdot (t_s + t_w \cdot m))$)
where $t_s$ is the communication start-up time, $t_w$ is the per-word transfer time,
and $m$ is the message size in words (alpha-beta cost model in the notation of \cite{grama}).
Using the reduction based on inverse recursive doubling (cf.\ Figure~\ref{fig:invdoubling}),
we obtain $T_C \in \Theta(\log p \cdot (t_s + t_w \cdot m))$ and, thus, $T_P \in \Theta(n/p + \log p \cdot (t_s + t_w \cdot m))$.
If our parameter $n$ is at least $p \cdot \log p$, then the {\em cost} $p T_p$, is in $\Theta(n)$
and we call our parallel algorithm \emph{cost optimal}. In other words, if each processing element gets to compute at least $\log p$ elements of the
sum, the overhead of the parallel version does not dominate the serial runtime asymptotically and therefore all processing elements can be used efficiently.%
\end{example}
\noindent
The above example also demonstrates the importance of efficient communication operations.
This is the focus of the next section, where we introduce our parallel data structures and
the computation and communication operations on them.

\section{Distributed Memory Parallel Data
Structures}
\label{sec:dpd}

At the heart of \framework lie the Distributed Memory Parallel Data Structures
(DPDs). The DPDs provide data-abstractions, but they rely on the user to
define partitions of the data to processing elements either directly through rank-data mappings or indirectly through lazy data wrappers.
The former case can be understood as a mapping of processing elements to the data they store locally. To understand the latter,
consider the situation where a user might define a two-dimensional grid representing a large matrix. Since this
structure would probably not fit in the memory available for a single processing element, each entry is wrapped in a lazy container.
Scala has language support for this and thus writing a lazy matrix-class can be achieved simply by adding the \textit{lazy} keyword
in front of the actual data field. In this way, each processing element will \emph{inflate} only the required matrices
at execution. While user-defined partitioning provides a lower level of abstraction than a dynamic workload balancing scheme would,
it provides a good balance between abstraction and analyzability. Alternatively, one of the available DPDs can
be used to conveniently map into data partitions, e.g.\ for a given array \verb!a!, we could use
\verb!(0 to a.length-1).toDistSeq.mapD(a)!, where \verb!a! is used as a partial function.

Once a DPD is instantiated, algorithms can be implemented directly through
chains of parallel transformations and communication methods directly on the
DPD. As usual in functional programming, in \framework,
DPDs are treated as immutable data structures and, consequently, all operations on DPDs
return new data structures.
\framework currently offers four DPDs: Distributed Values, Distributed
Sequences,
Distributed Grids, and Distributed Hash-Maps.
Due to the modularity of the communication operations in
\framework, the collections can easily be extended and expanded via user defined DPDs.
Such new distributed collections can either make use of the built-in group communications or
obtain direct peer-to-peer message passing capabilities (at the price of potential
deadlocks, etc.).
In this paper, we focus on the
Distributed Sequence DPD as the most common case. We also briefly introduce the Distributed Values DPD and the Distributed Grid DPD.

A constant or variable is the simplest form of a data structure, as it is
unstructured. \framework supports unstructured parallel data in the form of the
\textbf{Distributed Value DPD}. A Distributed Value is shared among all processing
elements of some communication group. As there is a local value for each processing element,
typical sequence operations like \verb!map!, \verb!reduce! etc.\ can
still be used, albeit without any sense of order or of sequence length. In other words, all processing
elements are participating in all operations and the reduction operators have to also be commutative

For Distributed Values, the processing element's rank is mapped to the local
data given as an argument at the point of instantiation. While this mapping is
trivial, it is still useful for operations that include all processing elements
in an execution.

\begin{example}
\label{ex:commonseed}
Agreeing on a dynamic random seed can be done in an unstructured way by taking the current unix time at
each node and agreeing and communicating the minimum of all these timestamps, thereby agreeing on a global seed.
{\normalfont
\begin{lstlisting}[language=scala]
val now: Long = System.currentTimeMillis
val min: Option[Long] = DistVal(now).allMinD
val rnd: Option[Random] = min.map(new Random(_))
\end{lstlisting}
}%
Here, \verb!allMinD! computes the minimum of the values in all processing elements and broadcasts it to all processing elements
in $\Theta(\log p)$. Alternatively, one could use the indexing method \verb!apply(0)! to get the current time from some arbitrary processing element with index $0$. It is even possible to share the random number generator itself. The reason for the type \verb!Option[Long]! is that for other DPDs, not all processing elements necessarily take part
in this computation. At the end, all processing elements have access to a random generator initialized with the global seed.
 We use the suffix \texttt{D} on parallel methods of DPDs in order to clearly distinguish between sequential and parallel operations. 
This is relevant especially since \verb!Option! supports similar higher order methods, e.g. \verb!dseq.reduceD(_+_).map(_+10)! would be a valid expression using a map on the \verb!Option! result.
\end{example}

The \textbf{Distributed Sequence DPD} distributes a sequence of a certain length to
processing elements. In other words, a distributed sequence is a one-to-one mapping of indices to ranks,
i.e., an index $i$ of the sequence is mapped to the numerical identifier of the processing element that stores the $i$-th element
of the sequence.

\begin{figure*}[t]
\centering
  \begin{tabular}{|l|l|l|}
  \hline
  \bf Operation&\bf Parallel Running Time $T_P$&\bf Communication\\\hline
  \verb!mapD! $\lambda$, \verb!foreach! $\lambda$&$\Theta(T_\lambda)$&None\\\hline
  \verb!apply! $i$&$\Theta((t_s + t_w \cdot m)\log p)$&Broadcast\\\hline
  \verb!reduceD! $\lambda$&$\Theta((t_s + t_w \cdot m + T_\lambda)\log p)$&Reduce\\\hline
  \verb!scan1D! $\lambda$&$\Theta((t_s + t_w \cdot m + T_\lambda)\log p)$&Prefix Sum\\\hline
  \verb!shiftD! $d$&$\Theta(t_s + t_w \cdot m)$&Circular Shift\\\hline
  \verb!sumD!, \verb!productD!, \verb!minD!, \verb!maxD!, \verb!avgD!&$\Theta((t_s + t_w \cdot m)\log p)$&Reduce\\\hline
  \verb!allSumD!, \verb!allProductD!, \verb!allMinD!, \verb!allMaxD!, \verb!allAvgD!&$\Theta((t_s + t_w \cdot m )\log p)$&All-Reduce\\\hline
  \end{tabular}
\caption{\it Table over operations supported by \framework DPDs. $m$ is message size in words, $p$ is \# of processing elements}
\label{fig:operations}
\end{figure*}

Such a rank-data mapping can be generated from any existing
indexed Scala sequence using \verb!toDistSeq! as exemplified in the first paragraph of this section
or
from a symbolic range such as \verb!(1 to n)! as seen in Example~\ref{ex:ppi}.
While creating distributed sequences from indexed sequences require each processing element
to contain the entire sequence, in combination with lazy
data elements, this restriction is lifted and actually turned into a powerful advantage. Consider for example a Distributed Sequence DPD
containing lazy references to matrices. The runtime overhead of dealing with this symbolic indexed sequence will be dominated
by the matrix-matrix multiplication used to reduce the sequence.

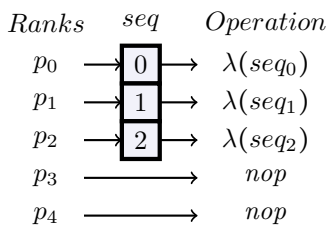
\begin{figure}[!Ht]
\centering
 \begin{tikzpicture}
    
    \def \s {0.5}
    \def \x {4}
    \def \n {2}
    \pgfmathtruncatemacro{\pn}{\n+2}
    \pgfmathtruncatemacro{\nplus}{\n+1}
    \pgfmathsetmacro{\shalf}{\s*0.5}

    \node at (\x-1,0.3) {$Ranks$};
    \node at (\x+\shalf,0.3) {$seq$};
    \node at (\x+\s+\s+0.9,0.3) {$Operation$};

    \foreach \y in { 0,...,\n} 
    {
      \draw [fill=lightgray,ultra thick] (\x,-\y*\s) rectangle
(\x+1*\s,-\y*\s-\s);
      \node at (\x+0.5*\s,-\y*\s-0.5*\s) {\y};
      
      \draw [->,thick] (\x-\s,-\y*\s-\shalf) -- (\x,-\y*\s-\shalf);
      \draw [->,thick] (\x+\s,-\y*\s-\shalf) -- (\x+\s+\s,-\y*\s-\shalf);
      \node at (\x+\s+\s+0.9,-\y*\s-\shalf) {$\lambda(seq_\y)$};
    }
    
    \foreach \y in { 0,...,\pn} 
    {
      \node at (\x-1,-\y*\s-0.5*\s) {$p_\y$};
    }
    
    \foreach \y in { \nplus,...,\pn} 
    {
      \draw [->,thick] (\x-\s,-\y*\s-\shalf) -- (\x+\s+\s,-\y*\s-\shalf);
      \node at (\x+\s+\s+0.9,-\y*\s-\shalf) {\textit{nop}};
    }
\end{tikzpicture}
\caption{\it Example of SIMD principle in mapD method invocation on Distributed Sequence of size 3.}
\label{fig:simd}
\end{figure}

Operations on the Distributed Sequence DPD are parallel versions of the typical sequence operations
such as \verb!map!, \verb!reduce! etc. Figure \ref{fig:simd} shows how the SIMD principle is mimicked in a parallel map operation. In contrast to the Distributed Values DPD, order and length of the sequence are respected and
only associativity is required for reduction operators.

The \textbf{Distributed Grid DPD} offers a more involved rank-data mapping than the other
DPDs, the motivation being a lack of efficient means for nested traversals of distributed
sequences representing multidimensional data. It offers an efficient rank-data mapping while providing convenience
methods for advanced communication patterns like subdimension partitioning. The mapping itself is a
generalization of positional notation for integers called \textit{mixed radix}
with the addition of a possible transposition. In a nutshell, integers can be converted to and from a list of subindices
for the individual dimensions. Section \ref{sec:floyd} shows how to use this to model a distributed matrix in a Floyd-Warshall parallelization.

\subsection{Group Communication Algorithms}

Central to the DPD operations of  \framework  are the \emph{group communication
algorithms}. These operations serve as the basis for \emph{all} network
communication in \framework,
but can be viewed as a completely encapsulated module. All the operations work
via \textit{asynchronous message passing}, i.e., nonblocking sends coupled with
blocking receives.
\framework introduces the notion of \texttt{FooPar Processes}, abstract processes
which encapsulate \textit{integer ranks} as well as \textit{send} and \textit{receive}
operations,  and, thus, can be treated as abstract processing
elements in group communication algorithms.

\framework adopts a topology-oblivious approach to the distribution of processing elements in groups, i.e., tasks are currently distributed in a round-robin fashion with no regard to cache coherence. Section~\ref{sec:empirical} shows that this is efficient in practice. However, 
\framework's design allows for future topology-aware heuristics to improve distribution of processing elements to optimize communication.

The group communication algorithms center around communication via message passing. Non-blocking sends and blocking receives are made possible through the process interface. 
We define the methods as  \verb!send(destination,! \verb!message)! and \verb!receive(origin)! where \verb!destination! and \verb!origin! are local ranks within a communication group.
This level of abstraction allows enough flexibility for a vast amount of algorithms to be implemented with no regard for actual network implementation or topology.

Arguably, the most trivial group communication operation is a \textbf{Circular Shift} by $d$ elements for a sequence of length $n$, where each process with rank $r$ sends its element to
$(r + d) \text{mod } n$  and receives an element from $(r - d) \text{mod } n$.

In Figure~\ref{fig:invdoubling} we have seen how to use inverse recursive doubling to implement the
\textbf{Reduction} group communication operation for associative reduction operators.
This operation is a special case of the higher-order \textit{fold} algorithm without an identity element.
The typical sequential fold comes in ordered left-to-right (\verb!foldLeft!), right-to-left (\verb!foldRight!) or unordered (\verb!fold!). To support these, data 
is required to be a \textit{monoid} structure, i.e., a semi-group with zero element. While the sequential fold imposes more restrictions on the structure of the data,
it offers more flexibility by supporting ordered folds on empty sequences as well as supporting arbitrary typed zero elements.
While left- and right-associative folds are inherently sequential, the reduction operation can be parallelized as discussed before. 

Finally, the elementary group communication \textbf{Broadcasting} which communicates a value from one
processing element to all others, can be performed by recursive doubling, i.e., the inverse of Figure~\ref{fig:invdoubling}, in
$\Theta(\log p)$.

To profit from the group communication patterns, high level data structure operations are mapped
to group operations. Scala sequences support second-order operations like for example \verb!map! and \verb!reduce!
as well as element retrieval using the indexing operator \verb!apply!.
Analogously, \framework offers variants of these on distributed sequences. The remainder of this section describes how and at what
(parallel) running time cost \verb!reduceD!, \verb!mapD!, and \verb!apply! can be implemented.

\noindent For performing computations on the local data of the processing elements, we do not need communication, but
we need to apply a function to each element of the distributed sequence.
A \texttt{mapD} with a function $\lambda$ of running time $T_\lambda$ can be performed independently in parallel,
yielding a parallel
running time of $\Theta(T_\lambda)$, as opposed to $\Theta(n \cdot T_\lambda)$
in the serial case. Here, we assume $n \leq p$, i.e., the existence of sufficiently many
processing elements. This assumption works because each element in the DPD can represent partitioned data.

The indexing operation, \texttt{apply}, needs communication. If each process needs a copy of the $i$-th element,
a Broadcast communication operation based on recursive doubling can be used to perform this operation in
$\Theta((t_s + t_w \cdot m) \log p)$ parallel
time. Here, $t_s$ and $t_w$ are the message startup time and per-word transfer time respectively.
Similarly,
\texttt{reduceD} can be performed in $\Theta((t_s + t_w \cdot m + T_\lambda)\log p)$ for
any reduction operation~$\lambda$.

For further operations, consult Figure~\ref{fig:operations}. There, two more communication operations are used.
All-Reduce is just like Reduce, except that all process elements receive the result of the reduction, while Prefix Sum
is a reduction where each processing element receives the partial reduction result up to its index. We use the
versions from \cite{grama}, except that we correct an obvious bug for Prefix Sum in order to avoid commutativity
as a requirement for the operation.

\begin{example}
The following code multiplies two matrices $A$ and $B$ represented by 2-dimensional Scala \texttt{Array}s of lazy matrix-wrappers
named \texttt{A} and \texttt{Bt}, where the latter represents $B^T$. Using \verb!mapD! and \verb!reduceD!, this runs in
parallel time 
$\Theta\left( \frac{n^3}{p}  +   \frac{n^2}{p^{2/3}}  \log p \right)$.
{\normalfont
\begin{lstlisting}[language=scala]
for (i <- 0 until M; j <- 0 until N)
 A(i) zip Bt(j) mapD {case (a,b) => a*b} reduceD (_ + _)
\end{lstlisting}
}%
\end{example}

\section{\framework Implementation in Scala}
\label{sec:implementation}

The mix between functional programming and object oriented features brings
interesting new possibilities for program design. In this section, the language concepts needed to
implement DPDs and their use in \framework are presented. In particular, our framework
builds on the {\em builder/traversable pattern} \cite{fsttcs2009} in order to obtain reusable and maintainable code by
reducing code duplication and boiler plate code as well as introducing a natural unification
of functional concepts. 

\subsection{Option monad}

Like Haskell, the standard Scala library also contains a \textit{maybe monad} \cite{wadler92}
structure in the shape of the parameterized \texttt{Option[+A]} trait. 
Simply put, a monad is an abstract concept that can be used for convenient control-flows in functional programming. 
\framework uses Scala's \texttt{Option} monad in order to enable the use of SIMD instructions for group operations where not all processing elements participate.

\framework also relies on the \texttt{Option} monad for robustness. In the battle against
\textit{NullPointerExceptions} known from \textit{Java}, Option makes the
possibility of a None value explicit, thus moving the problem to compile time. It is a huge
advantage for this framework to push as many errors as possible into compile time
to disallow erronous runs on expensive HPC-hardware with limited access to computation time.
One of the biggest advantages comes from the way Option supports SIMD
operations. %
If we consider the method \texttt{map}, \texttt{nop} instructions can be simulated elegantly.
From the definition of \texttt{Option} in the \textit{Scala} API 2.9.2 we have:
\pagebreak

\begin{lstlisting}[language=scala]
@inline final def map[B](f: A => B): Option[B] =
  if (isEmpty) None else Some(f(this.get))
\end{lstlisting}

\noindent This nicely encapsulates \texttt{nop} as a special case of the map operation on \texttt{Option} elements.
Given a list \texttt{xs} of options, we can nest a map operation to simulate \texttt{SIMD}:
\begin{lstlisting}[language=scala]
val xs = (1 to 4).map(i => if (i > 2) Some(i) else None)
def simd[T,U](f: T => U) = (o:Option[T]) => o.map(f)
def double(i:Int) = i*2
xs.map(simd(double))   // Vector(None,None,Some(6),Some(8))
\end{lstlisting}
Note that all the above code is sequential. However, \framework generalizes this pattern to parallel map operations supporting arbitrary SIMD operations in the form of $\lambda$ expressions.
Lists in Scala are also monads, the difference being that a list can be empty (\texttt{Nil}), or contain \textit{one or more} elements where Option supports none or \textit{exactly one} element.

To break it down, \framework uses options mainly in 2 cases: \textbf{1)} When not all processing elements participate in a group operation, the SIMD model enforces that they still make the method invocation, 
safely returning the type \texttt{Option[T]}, and \textbf{2)} when constructing a DPD, every processing element creates a symbolic structure with their local part stored in an \texttt{Option[T]} value, i.e., 
processing elements that are not a part of the communication group contain the \texttt{None} singleton but can continue to invoke methods safely.

\subsection{Implicits}
In Scala, implicits allow the compiler to choose appropriate values for expressions at compile time. Coupled with type-inference, this allows for advanced design patterns. 
\framework uses implicits extensively for its typeclasses and for the Traversable-Builder Pattern employed by its DPDs.
\begin{lstlisting}[caption=\it Generic base trait,label=lst:type]
trait A[T] {
  def filter(f: T => Boolean) = ??? //Need type B[T]
  def map[U](f: T => U) = ??? //Need type B[U]
}
class B[U](val x: U) extends A[U]
\end{lstlisting}
Consider the problem depicted in Listing \ref{lst:type} : Given two classes, \texttt{A[T]} and its subclass, \texttt{B[U]}, how can a method of \texttt{A[T]} have \texttt{B[U]} as a return type? 
This problem can be solved through implicit parameter resolution. If we define a generic method (\texttt{map}, Line 3 in Listing \ref{lst:builder}) we allow the compiler to choose
the generic type parameter based on available implicit values in companion objects (\texttt{builder}, Line 8 in Listing \ref{lst:builder}). Now, super-classes can implicitly work 
with sub-types through their definition of implicit builders. In turn, this allows for generic implementations of functions that return concrete types, i.e., \texttt{map[U,That]} is 
implemented with no knowledge of subclasses A and B, however, if we map an \texttt{A} into a container of \texttt{String}, we get a B.
\begin{lstlisting}[caption=\it Simple builder pattern,label=lst:builder]
trait Cont[T] {
  def elem: T
  def map[U, That](f: T => U)(implicit builder: U => That): That = builder(f(elem)) 
}
case class A(elem: Int) extends Cont[Int]
case class B(elem: String) extends Cont[String]
object B { //Companion object for B
  implicit def bldr:(String => B) = (s:String) => B(s)
}
object Impl extends App {
  val a:A = A(42)
  val b:B = a.map(x => x.toString)
}
\end{lstlisting}

\subsection{Type-Classes through Implicits}
\label{typeclasses}
\framework uses type-classes to provide numeric distributed operations on the DPDs, e.g. \texttt{sum, product, max, minBy} etc. Type-classes are flexible, and thus, \framework 
uses the \texttt{Numeric} type-class directly from the standard library of Scala, making numeric operations available for all implementing classes of the standard library. \framework goes 
even further and offers methods like \texttt{average} which are unavailable in the standard library.

A type-class \cite{hall1996} defines features for a set of types in a weaker sense than
interfaces or traits. In Scala, type-classes are powered by implicit arguments
to methods providing a looser coupling between classes than other
interface-constructs. Using the type-class pattern, a class can impose
constraints on a generic type, \texttt{T}, on a per-method basis rather than a
per-class basis as provided by interfaces and type-bounds. This concept is
called \textit{context bounds} in Scala, for example:


\begin{lstlisting}
def sumL[T:Numeric](l:List[T]) = l.sum
\end{lstlisting}
\noindent This unfolds to the following more verbose definition:
\begin{lstlisting}
def sumL[T](l:List[T])(implicit num:Numeric[T]) = l.sum
\end{lstlisting}

\noindent
Now we have a function definition which works only for types \texttt{T} which
provide an implicit numeric parameter, usually supplied by the companion object
for \texttt{T}. To explain how this is different than using an
\textit{upper type bound} directly on \texttt{T}, consider the following example:

\begin{lstlisting}
case class NumList[T <: Complex](xs: Complex*) {
  def sum = (xs fold new Complex(0, 0))(_ + _)
  def map[U <: Complex](f: Complex => U): NumList[U] = NumList(xs.map(f): _*)
  override def toString = "[" + xs.mkString(", ") + "]"
}
\end{lstlisting}

\noindent We define a numeric list based on lists of type \texttt{T} with upper bound
\texttt{Complex}. In this way, \texttt{NumList} gets access to numeric
operations like summation over the list. As a result, the list can be used for example
like this:

\begin{lstlisting}
val r = new Real(2)
val n = new Natural(10)
val comps = NumList(r, n, r, n)
println(comps)
println("sum: " + comps.sum)
println("sum * 2: " + comps.map(x => x + x).sum)
\end{lstlisting}
While this provides added functionality for numeric types, it comes at the
expense of an additional class definition and a bound on \texttt{T}. Furthermore, we were forced to
provide a template definition of addition (i.e., the method from
\texttt{Complex}). By introducing a \texttt{numeric} type-class, we unify the
concept of a generic list and a numeric list. With a numeric type-class, the
list class can be rewritten as follows:

\begin{lstlisting}
case class GenList[T](xs: T*) {
  def sum(implicit num: Numeric[T]) = xs.sum
  def map[U](f: T => U) = GenList(xs.map(f): _*)
  override def toString = "[" + xs.mkString(", ") + "]"
}
\end{lstlisting}
Now the list can be used for any type \texttt{T} while sum works for any type
\texttt{T} providing implicit definitions of \texttt{Numeric[T]}. In Scala,
many of the subclasses of \texttt{AnyVal} support this operation, so we have
successfully provided a numeric operation with a \textit{least knowledge}
principle about the generic type \texttt{T}.
Type-classes allow us to
use the smallest set of constraints for a type and provide us with very
fine-grained control.

Using implicit classes in combination with type-classes, we can extend existing
implementations of classes without modifications or additional interfaces. As
an example, we can add the \texttt{average} method to sequences of the Scala
API:

\begin{lstlisting}
implicit class AvgSeq[T](s: Seq[T]) {
def average(implicit num: Numeric[T]) = 
  num.toDouble(s.sum) / s.size
}
println(1 to 9 average)  //result: 5.0
\end{lstlisting}

\noindent We have seen a use of this already in Section~\ref{sec:dpd}, where a method
\texttt{toDistSeq} was added to Scala sequences.

Furthermore, \framework makes use of typeclasses to provide convenience methods for numeric types. As an example, consider the following extension to a Matrix class which can be considered a numeric type. 
We implement a subclass of the generic \texttt{Numeric[T]} trait for the type \texttt{Matrix}. 

\begin{lstlisting}[language=scala]
class MatrixIsNumeric extends Numeric[Matrix] with Ordering[Matrix] {
  def plus(x: Matrix, y: Matrix): Matrix = x + y
  def times(x: Matrix, y: Matrix): Matrix = x * y
  def negate(x: Matrix): Matrix = x * -1
...
}
\end{lstlisting}

\noindent By making an instance of this class implicitly available for the \texttt{Matrix} companion object, it will be accessible at compiletime for the \framework DPD classes.

\begin{example}
As an example, consider the 
reduction of a distributed sequence containing matrices of equal dimensions. By introducing the implicit numeric instance for matrices, the distributed sequence can be reduced via summation without 
imposing any constraints on the remaining methods of the distributed sequence.

{\normalfont
\begin{lstlisting}[language=scala]
val x = new Matrix(Seq(Seq(1, 2), Seq(3, 4)))
val dSeq = Array.fill(size)(x).toDistSeq
for (res <- dSeq.sumD) {
  pprintln(res, " size = " + size, res == x * size)
}
\end{lstlisting}
}

Line 1 creates an instance of the matrix class. Line 2 creates an array of processing elements in this \framework execution and then converts it to a distributed sequence. Line 3 calls the distributed {\normalfont\texttt{sumD}} method, which requires
an implicit instance of the {\normalfont\texttt{Numeric[Matrix]}} class. Using the for-comprehension, only the root process 
prints and checks the result against $x \cdot size$. Note that the for-comprehension can be used because 
Scala's {\normalfont\texttt{Option}} monad supports the higher order method {\normalfont\texttt{foreach}}. 

Running the above \framework program with 8 processing elements yields the following output as expected:\\[2ex]
\verb!Rank0: ([8.0,16.0]!\\
\verb![24.0,32.0], size = 8,true)!\\
\end{example}

\subsection{Builder/Traversable Pattern}
\label{sec:btp}
For \framework's implementation, a main goal was to avoid
reimplementing the distributed operations such as \texttt{reduceD} and \texttt{mapD}
for each DPD. With the \textit{Builder/Traversable} pattern we achieve this and implement them once and for all,
while maintaining the specific types of the DPDs.

One of the main benefits of Scala's implicits comes from the way the compiler
resolves implicit arguments at compiletime. 
\textit{``If there are several eligible arguments which match the implicit
parameter's type, a most specific one will be chosen using the rules of static
overloading resolution''} \cite{scalref}. Using this feature, type information
can be pushed \textit{upwards} in a type hierarchy via generics and an
adaptation of the \textit{Factory Pattern} \cite{fsttcs2009}, i.e., the
\textit{Builder/Traversable Pattern}, effectively solving the problem depicted in Listing \ref{lst:type}.

\begin{lstlisting}[language=scala]
trait DistTraversable[+T] {
  def elem: Option[T]
  def foreach(f: T => Unit): Unit
  def group: FooParGroup
}
\end{lstlisting}

\begin{figure*}
\centering
\includegraphics[scale=0.27]{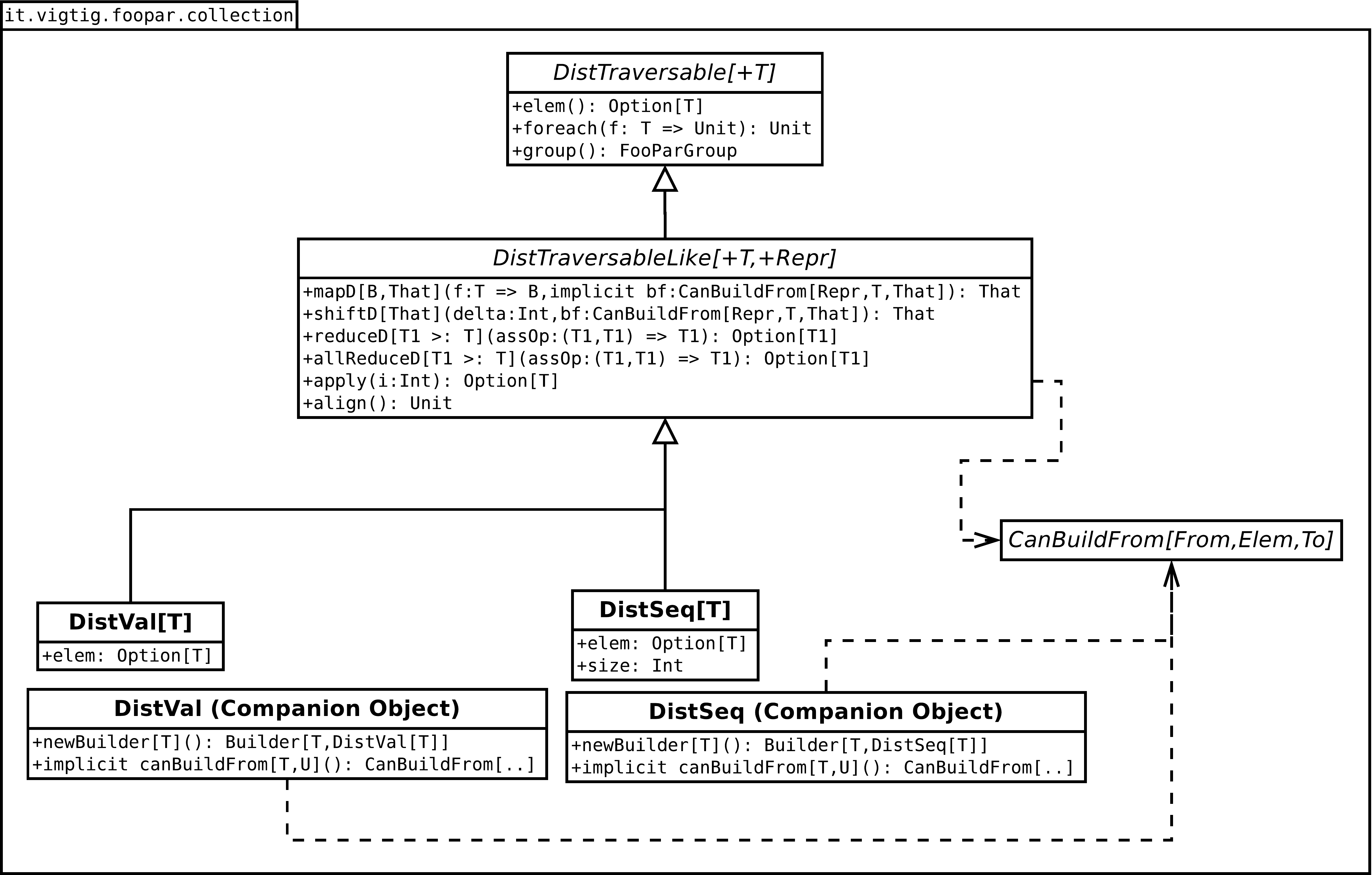}
\caption{\it Simplified UML diagram of collection package architecture. The
companion objects of \texttt{DistSeq} and \texttt{DistVal} provide type
information to \texttt{DistTraversableLike} through a loose coupling with
implicit definitions of \texttt{CanBuildFrom}.}
\label{fig:simplified}
\end{figure*}

\noindent The \texttt{DistTraversable} trait presented above
resides at the
base of the type hierarchy (a simplified overview can be seen in
Figure \ref{fig:simplified}).
It defines methods for retrieving a process' local element as well as means of
traversing it. The group is used to supply network communication operations and
follows a DPD through chains of consecutive transformations.

\sloppy{
\framework utilizes Scala's standard library support for the \texttt{Numeric[T]}
type-class \cite{hall1996}.} In combination with the group
communication operations the user gains convenient access to distributed
versions of operations like \texttt{sum, average, min, max} and \texttt{product}
for types \texttt{T}, which provide implicit \texttt{Numeric[T]} values. This
allows user-defined
algebraic types (as well as standard primitives) to work with the built in DPDs in
\framework. The distributed numerical operations are available through
the \texttt{TraversableNumOps} trait.

\noindent Both for the type-class support of numeric operations as well as for applying
the \textit{Traversable/Builder} pattern, \framework makes heavy use of
\textit{context bounds}.

In the following we are going to take a look at the implementation of
distributed variables. The implementation of e.g.\ distributed sequences
is analogous except for more complex builders due to the mapping of
indices to ranks.

As shown below \texttt{DistVal[+T]} is basically just the
composition of the traits \texttt{DistTraversableLike} and
\texttt{TraversableNumOps} and thus the most basic implementation of a
DPD.

\begin{lstlisting}[language=scala]
class DistVal[+T]
(val elem: Option[T], val group: FooParGroup)
(implicit val fpapp: FooParApp)
extends DistTraversableLike[T, DistVal[T]]
  with TraversableNumOps[T, DistVal[T]] {
  def size = 1
}
\end{lstlisting}
The real work is performed in \texttt{DistTraversableLike} as well as in the implicit builders of the companion object:
\begin{lstlisting}[language=scala]
object DistVal {
  type DTB[T] = DistTraversableBuilder[T]
  implicit def canBuildFrom[T, U]
  (implicit fpapp: FooParApp): CBF[DistVal[T], U, DistVal[U]] =
    new CBF[DistVal[T], U, DistVal[U]] {
      def apply(): Builder[U, DistVal[U]] = newBuilder(None)
      def apply(from: DistVal[T]): Builder[U, DistVal[U]] =
        newBuilder(Some(from.group))
    } ...
}
\end{lstlisting}

\noindent\sloppy
When the companion objects of an extending class supplies an implicit builder
through the generic trait \texttt{CanBuildFrom[From,Elem,To]}, super classes 
can access this builder at compile-time using explicit self-typing.
The companion object \texttt{DistVal} implements such an implicit definition
of the \texttt{CanBuildFrom} trait.

\sloppy
Finally, the trait \texttt{DistTraversableLike[T,Repr]} defines the collection operations
associated with a distributed traversable by using an implicit builder as a
parameter to the respective methods as seen in the below code.


%
\begin{lstlisting}[language=scala]
trait DistTraversableLike[+T, +Repr] extends DistTraversable[T] {
  self: Repr =>
  ...
  def reduceD[T1 >: T](assOp: (T1, T1) => T1): Option[T1] = {
    if (group.partOfGroup)
      for (x <- this) group.allOneReduce(x, assOp).foreach(y => return Some(y))
    return None
  }

  def mapD[B, That](f: T => B)(implicit bf: CanBuildFrom[Repr, B, That]): That =
  {
    val b: Builder[B, That] = bf(this)
    if (group.partOfGroup)
      for (x <- this) b += f(x)
    b.result
  } ...
}
\end{lstlisting}
As shown in Figure \ref{fig:simplified}, the same approach is used for
distributed sequences (and other DPDs supported by \framework).

\section{Verification and Testing}
\label{sec:verifytest}

Using core features of functional programming like \textit{higher-order functions} and \textit{immutable data structures}, \framework allows for analyzable and verifiable programs 
to be designed and implemented. Given a higher order function, $f$, which takes a function value parameter, $\lambda$, one can prove the correctness of $f$ given some properties of $\lambda$.
Furthermore, higher order functions can abstract away message passing, including deadlock-safety in their proofs. As an example, \framework provides a distributed method, \texttt{reduceD}$(\lambda)$, which is proven to be 
correct and deadlock-safe for any associative binary operator, $\lambda$. Once a program can be modeled purely in terms of proven higher-order 
functions, the correctness and safety of that program follows from properties of the function-parameters. In this section, we give a short proof of \textit{All-to-One Reduction} and show 
how specification testing can be used to bridge the gap between proof and practice.
\subsection{Correctness Proof of All-to-One Reduction}

\begin{algorithm}
\caption{\it Functional All-to-One Reduction in \framework.\\[-1ex]}
\label{alg:reduce}
\small
\ \\[-5ex]
\begin{algorithmic}
\Function{reduce}{$\lambda,size,acc,rank$}
  \For{\textit{i} $\leftarrow  0$ \textbf{until} $\lceil\log_2 size\rceil - 1 $} 
    \State m $\leftarrow 2^i$
    \If {$rank \bmod 2m = 0$} 
      \State \textit{acc} $\leftarrow \lambda(acc,$ rcv$(rank+m))$ \Comment{Side-effect}
    \ElsIf { $rank \bmod m = 0$ }
      \State send($rank-m,acc$) \Comment{Side-effect}
    \EndIf
  \EndFor
  
  \If {$rank = 0$} 
    \Return Some(value)
  \Else \ \Return None
  \EndIf
\EndFunction
\end{algorithmic}\ \\[-6ex]
\end{algorithm}

\noindent
We prove the correctness of the All-to-One Reduction in \framework (Algorithm \ref{alg:reduce}) for powers of two by induction on the number of processing elements $p$. Without loss of generality, we use the associative operator $\lambda := (\cdot)$ and $rank = 0$ as root element. The case for $p=1$ is trivial in that no communication takes place. For illustration purposes, we choose $p=4$ as the base-case instead.

\textbf{Base step:} Let $p = 2^n, n=2$ and $e_r$ the element
initially contained by processing element $r$.
\begin{center}
\begin{tabular}{|l|l|l|l|c|}\hline
$i$ & $m$ & $rank$ & recieve/send & $acc$\\\hline
0& 1 & $0$ & $receive(1)$ & $(e_0\cdot e_1)$\\
 && $1$ & $send(0,e_1)$ & $e_1$\\
 && $2$ & $receive(3)$ & $(e_2\cdot e_3)$\\
 && $3$ & $send(2,e_3)$ & $e_3$\\\hline
1& 2 & $0$ & $receive(2)$ & $((e_0\cdot e_1)\cdot(e_2\cdot e_3))$\\
&& $1$ & $nop$ & $e_1$\\
& & $2$ & $send(0,(e_2\cdot e_3))$ & $(e_2\cdot e_3)$\\
&& $3$ & $nop$ & $e_3$\\\hline
\end{tabular} 
\end{center}
We see that after 2 iterations, $rank = 0$ contains $\prod_{r=0}^{p-1}e_r$, concluding the base step. Let the induction hypothesis state that, for $p=2^n$, the processing element with $rank = 0$ will contain $\prod_{r=0}^{2^n-1}e_r$ after the $n$th iteration.

\textbf{Induction step $(n+1)$:} If $p=2^{n+1}$, at the beginning of iteration $i=n$, by the induction hypothesis, the first $2^n$ elements complete the partial reduction at the root node, $rank = 0$. Furthermore,
we see that the remaining $2^n$ processing elements of the \textit{upper half} complete the analogous algorithm at $rank = 2^n$, computing $\prod_{r=2^n}^{2^{n+1} -1}e_r$. At iteration $i=n$, $m=2^n$ and thus, 
$rank = 0$ receives the reduction of the upper half from $rank = {2^n}$. Since $$\prod_{r=0}^{2^{n+1}-1}e_r = \left(\prod_{r=0}^{2^n -1}e_r\right) \cdot \left(\prod_{r=2^n}^{2^{n+1}-1}e_r \right)$$ $rank = 0$ now contains the 
reduction of the entire sequence $\Box$.

To prove that this algorithm is safe, we follow the same pattern as the correctness proof. To go from the case of $p=2^n$ to the case $p=2^{n+1}$, we complete the final step of the reduction by sending a message from $rank = {2^n}$ to $rank = 0$, unlocking the resource necessary for 
$rank = 0$ to complete its receive call.

Note that the order of computation respects associativity and thus works for any binary associative function. By the fact that Recursive Doubling is the exact inverse operation of All-to-One Reduction using an appropriate $\lambda$
function, we conclude that both the correctness proof and the safety property hold for Recursive Doubling. Finally, the algorithm and proof can (with only minor additions) be generalized to 
arbitrary roots \footnote{Reduction to other roots can obviously be achieved by xor-ing all ranks in the algorithm with the rank of the target root.} and arbitrary number of processing elements.

\subsection{Testing Properties}
While the combination of pseudo-code, correctness/safety proofs, and parallel asymptotic running time and scalability analysis can take us far, it is useless if the 
theory and implementation remain disconnected. Even in a high-level language like Scala, it can be extremely tough to show that the implementation adheres to a strict 
contract between the pseudo-implementation and the code. Specification testing \cite{odersky10} is a natural next step in Unit Testing. At its core, it is an abstraction over 
test-case construction, using generators for core types that can be extended to support new data structures. Using \textit{ScalaCheck}, \framework provides a high level of code-trust, 
solidifying the relationship between proofs and properties found theoretically and the actual code. Since Scala supports imperative programming, detailed pseudo-code can be translated 
directly into program code. If imperative code pertains to inherently stateful parts of a program (e.g. message passing in group communication patterns), and is interfaced only through stateless functional programming, 
the disadvantages of imperative programming do not bleed into user programs. In addition, we can modularize proofs elegantly through proper use of higher order functions and immutability.


\begin{lstlisting}[label=lst:spec,caption=\it Specification Testing in FooPar]
property("maxD") = forAll(sizedLists) { xs: List[Int] =>
  val dseq = xs.toDistSeq
  dseq.maxD shouldEqual xs.max
}
property("reduceD with concat") = forAll(sizedLists) { xs: List[Int] =>
  val dseq = xs.toDistSeq.mapD(_.toString)
  dseq reduceD (_+_) shouldEqual xs.mkString
}
\end{lstlisting}

Listing \ref{lst:spec} shows how specification testing can programmatically provide arbitrary test-cases (i.e., automatic edge-case construction) without 
huge amounts of boiler-plate code. The \texttt{shouldEqual} method is a small \framework extension that allows for idiomatic use of the option monads returned from 
the DPD methods.\\[-4ex]

\begin{lstlisting}[label=lst:should,caption=\it Implicit extension with shouldEqual method]
implicit class Should[T](opt: Option[T]) {
  def shouldEqual[U](x: U) = opt.map(_ == x).getOrElse(true)
}
\end{lstlisting}
Listing \ref{lst:should} shows how the method \texttt{shouldEqual} utilizes the \texttt{Option} monad to provide a fallback making sure that processing elements that do not
receive a result always answer \texttt{true} to specification testing.

\section{Empirical Results}
\label{sec:empirical}
 
 \begin{figure}[t]
 \centering
 \begin{tikzpicture}[scale=1]
\begin{axis}[
height=6cm,
width=8cm,
legend style={
cells={anchor=east},
at={(0.5,1.2)},anchor=north,
legend plot pos=right
},
xlabel={$n$},
ylabel={seconds},
legend columns=2,
ymax=21, ymin=0,
xmin=100, xmax=600,
grid=major,
symbolic x coords={100,200,300,400,500, 550, 600},
cycle multi list={
{red,mark=square*},
{blue,mark=*}
}]
\addplot coordinates {
(100,0.015295900000000001)
(200,0.0729571)
(300,0.2477926)
(400,0.7547429)
(500,1.504653)
(550,3.7086813999999997)
(600,5.6438586)
};

\addplot coordinates {
(100, 0.0375829)
(200, 0.27488579999999996)
(300, 0.9140756)
(400, 2.7307048000000003)
(500, 5.2165314)
(550, 12.7171593)
(600, 20.0928842)
}; 

\legend{$FooPar_{Akka}$,$Java_{MPJE}$}

\end{axis}
\end{tikzpicture}
\caption{\it Average walltime in seconds for matrix reduction with multiplication.
\framework with Akka backend and Java with MPJ Express on 16 cores spread across
2 machines.} 
\label{fig:reducetest}
\end{figure}
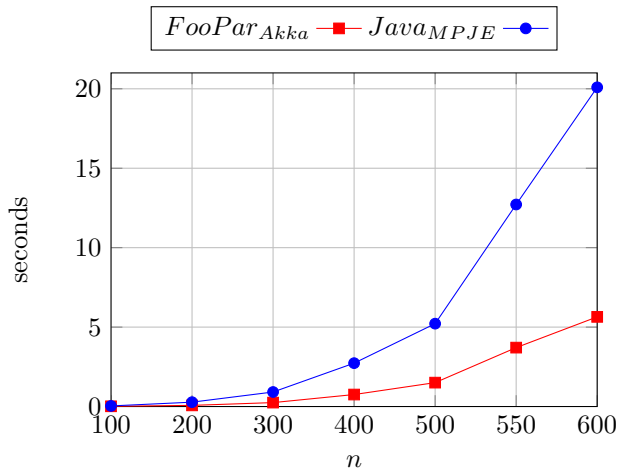
\framework's ability to reach close-to-optimal performance in real world HPC settings has already been demonstrated extensively for dense matrix-matrix multiplication (cf.\ Figure~\ref{fig:efficiency}) in \cite{ppam13}. Thus, in this section, we focus on two other aspects. First, we compare the performance of a \framework reduction to that of the MPJ Express implementation. Second, we show the framework's ability to scale by presenting 
a parallel implementaiton of the Floyd-Warshall algorithm in \framework and comparing it to a sequential implementation.

\subsection{Comparison to MPJ Express}

We conducted a test of computational scalability reducing $p$ matrices of
varying size with the associative operation of matrix-matrix multiplication. By
varying the size of the matrices we show the performance impact of the
computation part in a reduction in \framework compared to Java+MPJ Express.
Both use the same na\"{\i}ve Java implementation of matrix multiplication utilizing
a 1-dimensional \texttt{double} array matrix representation.

\sloppy
An analysis of the source of MPJ Express reveals an inefficient implementation of the reduce
operation. For matrices of size $n$, it runs in ${T_P \in \Theta((p-1)(T_s+T_w(n^2)+n^3))}$, whereas
\framework \linebreak implements this operation in ${T_P \in \Theta(n( \log p
(T_s+T_w(n^2)}$ ${+n^3) )}$. Note that for these running times we assume sufficient bandwidth between
nodes at each stage of the respective algorithms, i.e., they are network
topology oblivious \cite{grama}.

Even when using MPJ Express as a backend,
\framework can still achieve the latter parallel running time because it adds an
algorithmic layer for group communication operations directly above message
passing, thereby diminishing the relevance of backend supplied group
operations. 
Figure \ref{fig:reducetest} shows the expected tendency of running times. Notice
that, even for small sizes of $n$, \framework dominates Java with MPJ Express.

\subsection{Floyd-Warshall Parallelization}
\label{sec:floyd}
The Floyd-Warshall algorithm solves the all-pairs shortest path problem between
all nodes $v_1, \ldots, v_n$ in a weighted graph. Let $d_{i,j}^k$ be the weight
of the \textit{minimum-weight path} between $v_i$ and $v_j$ among vertices in
the set $\{v_1,\ldots,v_k\}$. The weight of an edge between nod $v_i$ and $v_j$ is denoted as $w(v_i,v_j)$.

The dynamic programming formulation can be
expressed as follows, where the shortest path from $v_i$ to $v_j$ is given by $d_{i,j}^n$:\\[1ex]

$
 d_{i,j}^k = \begin{cases}
              w(v_i,v_j) &, k = 0\\
	      \min \left\{ d_{i,j}^{k-1}, d_{i,k}^{(k-1)} + d_{k,j}^{(k-1)}\right\}  &, k \geq 1
             \end{cases}
$\\[2ex]

\begin{figure}
\begin{center}%
\begin{tikzpicture}[scale=0.8]

\def\n{8}
\def\u{1}
\def\half{\u*0.5}
\def\third{\u/3}
\def\fourth{\u/4}

\def\crossx{2}
\def\crossy{5}

\draw[help lines,thick] (0,0) grid (\n,\n);

\foreach \x in {0,1,...,7} {

	\draw [fill=gray!10] (\crossx,\x) rectangle (\crossx+1,\x+1);
	\draw [fill=gray!10] (\x,\crossy) rectangle (\x+1,\crossy+1);

}

\node[above,red,font=\bfseries] at (\crossx+\half,\n+\third) {$k$ column};
\draw[-,red,] (\crossx+\half,\n) -- (\crossx+\half,0);

\node[left,red,font=\bfseries] at (0-\third,\crossy+\half) {$k$ row};
\draw[-,red] (0,\crossy+\half) -- (\n,\crossy+\half);

\foreach \x in {0,1,...,7} {

	\ifthenelse{\NOT \x = 1}{
	\draw[->,thick] (\crossx+\half+\third,\x+\half) -- (\crossx+1+\half,\x+\half);
	}{};
	\draw[<-,thick] (\crossx-\half,\x+\half) -- (\crossx+\half-\third,\x+\half);

	\draw[->,thick] (\x+\half,\crossy+\half+\third) -- (\x+\half,\crossy+1+\third);
	\ifthenelse{\NOT \x = 6}{
	  \draw[->,thick] (\x+\half,\crossy+\half-\third) -- (\x+\half,\crossy-\third);
	}{};
}

\node[fill=white,circle,draw=none,inner sep=0pt,minimum size=21pt] at (6+\half+\fourth,1+\half+\fourth) {$d_{l,r}^{k}$};
\draw [fill](6+\half,1+\half)  circle [radius=0.1];

\node[fill=white,circle,draw=none,inner sep=0pt,minimum size=15pt] at (2+\fourth,2-0.15) {$d_{l,k}^{k-1}$};
\draw [fill](2+\half,1+\half)  circle [radius=0.1];

\node[fill=white,circle,draw=none,inner sep=0pt,minimum size=21pt] at (6+\half+\third,5+\half+\fourth) {$d_{k,r}^{k-1}$};
\draw [fill](6+\half,5+\half)  circle [radius=0.1];

\draw[->,thick,dashed]  (\crossx+\half+\third,2-\half) -- (6-\third,2-\half);
\draw[->,thick,dashed]  (6+\half,\crossy+\half-\third) -- (6+\half,2+\third);

\end{tikzpicture}\end{center}
\caption{\it Communication pattern employed in
parallel
Floyd-Warshall:
in iteration $k$, messages of size $n/\sqrt{p}$ are broadcast along $\sqrt{p}$ processing elements (in rows and columns).}

\label{fig:floyd}
\end{figure}
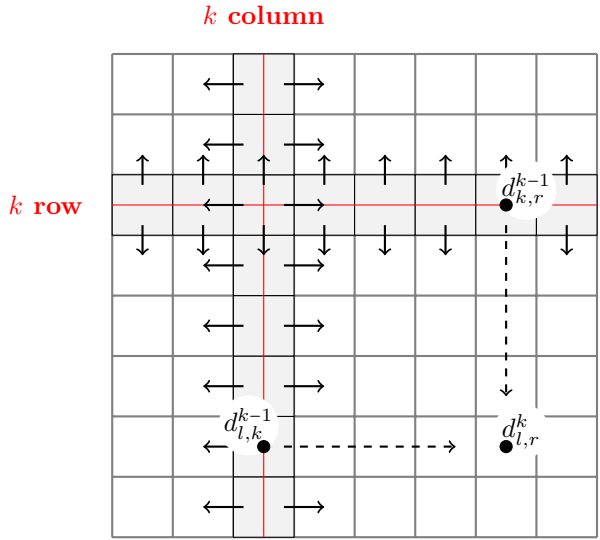

\noindent We follow
the parallelization approach from \cite{floyd} (communication pattern presented in Figure \ref{fig:floyd}) and present a scalable version of
the \linebreak parallel Floyd-Warshall algorithm that employs \framework's Distri\-buted Grid (size
$p=q^2$, i.e., $q$ processing elements per dimension):


%
\begin{lstlisting}[caption=\it Floyd-Warshall implementation in FooPar,label=lst:fw]
def update(row: Vector, col: Vector)(mat: Matrix) = {
  for (i <- 0 until mat.size; j <- 0 until mat(0).size)
    mat(i)(j) = math.min(mat(i)(j), row(j) + col(i))
  mat
}
def floyd(blocks: LazyMatrix, BS: Int) = {
  val dim = blocks.size
  val R = 0 until dim
  val N = dim * BS
  var grid = DistGrid(R, R) mapD { case i :: j :: Nil => blocks(i)(j).data }
  for (k <- 0 until N) {
    val ik = grid.ys.mapD(_(k % BS)).apply(k / BS).get
    val kj = grid.xs.mapD(_.map(_(k % BS))).apply(k / BS).get
    grid = grid.mapD(update(ik, kj))
  }
  grid
}
\end{lstlisting}

\noindent Briefly explained, Lines 7-10 initialize the 2-dimensional grid and Line 12 is
the inherent sequential loop of the algorithm, which is safely modeled as
a standard for loop. 
Line 13 gets the row $(k \bmod BS)$ of block $\lfloor k / BS \rfloor$ in
the \textit{column} of the calling process.
Similarly, Line 14 gets the column $(k \bmod BS)$ of the block $\lfloor k / BS
\rfloor$ in the \textit{row} of the calling process.
Line 15 transforms the grid into the next iteration by updating each block in parallel.

 \begin{figure}[t]
\centering
\begin{tabular}{|l|l||l|l||l|l|}\hline
 \textbf{n} & \textbf{p}& \multicolumn{2}{|c||}{\textbf{Speedup}}& \multicolumn{2}{|c|}{\textbf{Running time in seconds}} \\ %
 & &  MPJ & Akka & MPJ & Akka \\\hline
 \multirow{7}{*}{10080} & 16 &14.65 & 12.76&408.52 &469.06 \\ 
  & 25  &22.56 &19.93&265.3 &300.33 \\ 
  & 36  &26.86&27.40 &222.85 &218.5 \\ 
  & 49  &39.07& 33.81&151.25 &177.05 \\ 
  & 64  &46.17& 36.81&129.66 &162.63 \\ 
  & 81  &56.42&45.07&106.09 &132.83 \\ 
  & 100  &63.54& 60.55&94.21 &98.86 \\ \hline\hline
  \textbf{38000}& \textbf{100} & \textbf{94.28} & 87.39 & 3401.68 & 3669.86 \\ \hline
\end{tabular}

\caption{\it Floyd-Warshall parallel benchmark with input size $n=10080$  and varying numbers of processing elements $p$.
The row for $n=38000$ and $p=100$ exemplifies speedups on large-scale inputs.} 
\label{fig:floydresults}
\end{figure}
Figure \ref{fig:floydresults} shows scalability result for this implementation. We reach efficiencies of $\approx0.94$ and $\approx0.87$ with MPJ-Express and Akka respectively,
i.e., we see that the algorithm is scalable even for large numbers of processing elements. Furthermore, while Akka is mostly dominated by MPJ-Express, the backends behave similarly and the difference in performance 
are likely caused by differences in constants like $t_s$ or $t_w$, i.e., startup and per-word transfer times. 

While this example showcases the power of the abstractions in \framework, no work has been done to optimize the computational kernel for the update function (Lines 1-5 of Listing \ref{lst:fw}). We note, however, that 
a highly rewarding aspect of using higher-order functions is that computational kernels are naturally separated from the overall algorithm and, more importantly, completely disconnected from the communication code.

\section{Conclusion}
\label{sec:conclusion}

We have presented \framework, a novel Scala framework for massively parallel distributed memory
computing in Scala, which allows for concise and elegant high-level formulations of
parallel algorithms by abstracting the underlying communication into high-level group
communication operations.

A \framework implementation of the Floyd-Warshall algorithm using distributed grids achieves a near-linear speed-up of
more than 94 on a cluster using 100 processing elements for a matrix of dimension $38000 \times 38000$ demonstrating \framework's potential
for expressing scalable algorithms, in this example reducing a computation of nearly 4 days to less than 1 hour.

We have also shown that, independently of the backend, \framework can achieve extremely competitive performance due
to its judicious implementation of the high-level abstractions based on the \textit{Builder/Traversable} pattern.

\subsection{Relation to Previous Work}

\framework was very recently introduced in \cite{ppam13}, wich focuses on parallel algorithms, isoefficiency analysis, and absolute performance. In contrast, in this paper we focus on its architecture and programming model, its implementation in Scala, its comparative performance w.r.t.\ MPJ Express, and its real-world scalability.

\subsection{Future Work}

First, we observe that workload partitioning represents a threshold point of abstraction for parallel programming.
This problem is less pronounced in shared memory architectures, as the partitioning of data can be symbolic. In
distributed memory settings, the communication cost of workload partitioning can easily become a bottleneck of an algorithm.
Furthermore, parallel programs using automated workload partitioning are often harder to analyze.
While automated workload partitioning does not fit directly into \framework,
we plan to explore how \framework can be integrated with a separate dynamic workload allocation module,
in particular for local shared memory workload partitioning in multi-core nodes.

Thirdly, \framework's mix of the SPMD and SIMD paradigms makes it ideal for use in connection with large scale SIMD hardware. CUDA \cite{cuda} and OpenCL \cite{opencl} offer interesting takes on SPMD/SIMD programming,
especially w.r.t. GPU based systems. GPU executions of DPD methods could offer
performance boosts to single-node operations,
in combination with the distributed algorithms already possible in \framework.

\section*{Acknowledgements}

We acknowledge the support by the Danish Council for Independent Research, the Innovation
Center Denmark, the Lawrence Berkeley National Laboratory, and the Scientific Discovery
through Advanced Computing (SciDAC) Outreach Center. 

\vspace{3ex}

\bibliographystyle{plain}

\bibliography{fhpc14}

\begin{thebibliography}{10}

\bibitem{urltocarver}
Carver.
\newblock \url{http://nersc.gov/users/computational-systems/carver/}.

\bibitem{Alle07a}
Eric Allen, David Chase, Joe Hallett, Victor Luchangco, Jan-Willem Maessen,
  Sukyoung Ryu, Guy~L. Steele~Jr., and Sam Tobin-Hochstadt.
\newblock {The Fortress Language Specification}.
\newblock Technical report, Sun Microsystems, Inc., 2007.

\bibitem{ctwatch06}
David~A. Bader, Kamesh Madduri, John~R. Gilbert, Viral Shah, Jeremy Kepner,
  Theresa Meuse, and Ashok Krishnamurthy.
\newblock Designing scalable synthetic compact applications for benchmarking
  high productivity computing systems.
\newblock {\em CTWatch Quarterly}, 2(48), November 2006.

\bibitem{rauchwerger10}
Antal Buss, Harshvardhan, Ioannis Papadopoulos, Olga Pearce, Timmie Smith,
  Gabriel Tanase, Nathan Thomas, Xiabing Xu, Mauro Bianco, Nancy~M. Amato, and
  Lawrence Rauchwerger.
\newblock Stapl: Standard template adaptive parallel library.
\newblock In {\em Proceedings of the 3rd Annual Haifa Experimental Systems
  Conference}, SYSTOR '10, pages 14:1--14:10, New York, NY, USA, 2010. ACM.

\bibitem{chapel}
B.~L. Chamberlain, D.~Callahan, and H.~P. Zima.
\newblock Parallel programmability and the chapel language.
\newblock {\em IJHPCA}, 21(3):291--312, 2007.

\bibitem{kumar09}
Nicholas Chen, Rajesh~Kumar Karmani, Amin Shali, Bor-Yiing Su, and Ralph
  Johnson.
\newblock Collective communication patterns.
\newblock In {\em ParaPLOP}, 2009.

\bibitem{pgas05}
Cristian Coarfa, Yuri Dotsenko, John Mellor-Crummey, Fran\c{c}ois Cantonnet,
  Tarek El-Ghazawi, Ashrujit Mohanti, Yiyi Yao, and Daniel
  Chavarr\'{\i}a-Miranda.
\newblock An evaluation of global address space languages: {C}o-{A}rray
  {F}ortran and {U}nified {P}arallel {C}.
\newblock In {\em PPoPP}, pages 36--47. ACM, 2005.

\bibitem{spmd}
F.~Darema.
\newblock The {S}{P}{M}{D} model : Past, present and future.
\newblock In {\em EuroPVM/MPI Conference}, LNCS 2131, page~1. Springer, 2001.

\bibitem{gabriel04}
Edgar Gabriel, Graham~E. Fagg, George Bosilca, Thara Angskun, Jack~J. Dongarra,
  Jeffrey~M. Squyres, Vishal Sahay, Prabhanjan Kambadur, Brian Barrett, Andrew
  Lumsdaine, Ralph~H. Castain, David~J. Daniel, Richard~L. Graham, and
  Timothy~S. Woodall.
\newblock Open {MPI}: Goals, concept, and design of a next generation {MPI}
  implementation.
\newblock In {\em European PVM/MPI Users' Group Meeting}, pages 97--104, 2004.

\bibitem{grama}
A.~Grama, G.~Karypis, V.~Kumar, and A.~Gupta.
\newblock {\em Introduction to Parallel Computing}.
\newblock Pearson, Addison Wesley, 2003.

\bibitem{hall1996}
Cordelia~V. Hall, Kevin Hammond, Simon~L. Peyton~Jones, and Philip~L. Wadler.
\newblock Type classes in haskell.
\newblock {\em ACM TOPLAS}, 18(2):109--138, 1996.

\bibitem{ppam13}
Felix~P. Hargreaves and Daniel Merkle.
\newblock {FooPar: A Functional Object Oriented Parallel Framework in Scala}.
\newblock In {\em PPAM}, number 8385 in LNCS, pages 118--129, 2014.
\newblock preprint:\url{http://arxiv.org/abs/1304.2550}.

\bibitem{opencl}
Kamran Karimi, Neil~G. Dickson, and Firas Hamze.
\newblock A performance comparison of {CUDA} and {OpenCL}.
\newblock {\em CoRR}, abs/1005.2581, 2010.

\bibitem{floyd}
V.~Kumar and V.~Singh.
\newblock Scalability of parallel algorithms for the all-pairs shortest path
  problem.
\newblock {\em JPDC}, 13(2):124--138, 1991.

\bibitem{eden05}
Rita Loogen, Yolanda Ortega-Mall\'{e}n, and R.~{Pe\~{n}a}.
\newblock Parallel functional programming in {E}den.
\newblock {\em JFP}, 15:431--475, 2005.

\bibitem{apgas07}
Ewing Lusk and Katherine Yelick.
\newblock Languages for high-productivity computing: the {DARPA HPCS} language
  project.
\newblock {\em PPL}, 89(17), 2007.

\bibitem{x10}
Josh Milthorpe, V.~Ganesh, Alistair Rendell, and David Grove.
\newblock X10 as a parallel language for scientific computation: Practice and
  experience.
\newblock {\em Parallel and Distributed Processing Symposium}, pages
  1080--1088, 2011.

\bibitem{cuda}
John Nickolls, Ian Buck, Michael Garland, and Kevin Skadron.
\newblock Scalable parallel programming with cuda.
\newblock {\em Queue}, 6(2):40--53, 2008.

\bibitem{fsttcs2009}
M.~Odersky and A.~Moors.
\newblock Fighting bit rot with types (experience report: Scala collections).
\newblock In {\em FSTTCS}, LIPIcs 4, pages 427--451, 2009.

\bibitem{odersky10}
Martin Odersky.
\newblock Contracts for {S}cala.
\newblock In {\em Runtime Verification (RV)}, LNCS 6418, pages 51--57.
  Springer, 2010.

\bibitem{scalref}
Martin Odersky.
\newblock The {S}cala language specification, 2011.

\bibitem{odersky11}
Aleksandar Prokopec, Phil Bagwell, Tiark Rompf, and Martin Odersky.
\newblock A generic parallel collection framework.
\newblock In {\em Euro-Par 2011 Parallel Processing}, LNCS 6853, pages
  136--147. Springer, 2011.

\bibitem{rb12}
R.~Roestenburg and R.~Bakker.
\newblock {\em Akka in Action}.
\newblock Manning, 2012.

\bibitem{rompf13}
Tiark Rompf, Arvind~K. Sujeeth, Nada Amin, Kevin~J. Brown, Vojin Jovanovic,
  HyoukJoong Lee, Manohar Jonnalagedda, Kunle Olukotun, and Martin Odersky.
\newblock Optimizing data structures in high-level programs: New directions for
  extensible compilers based on staging.
\newblock {\em SIGPLAN Not.}, 48(1):497--510, January 2013.

\bibitem{rompf11}
Tiark Rompf, Arvind~K. Sujeeth, HyoukJoong Lee, Kevin~J. Brown, Hassan Chafi,
  Martin Odersky, and Kunle Olukotun.
\newblock Building-blocks for performance oriented dsls.
\newblock In Olivier Danvy and Chung chieh Shan, editors, {\em DSL}, volume~66
  of {\em EPTCS}, pages 93--117, 2011.

\bibitem{shafi09}
A.~Shafi and J.~Manzoor.
\newblock Towards efficient shared memory communications in {M}{P}{J} express.
\newblock In {\em IPDPS}, pages 1--7, 2009.

\bibitem{taboada_jos10}
G.~L. Taboada, J.~Touri\~no, and R.~Doallo.
\newblock {F-{M}{P}{J}: Scalable Java Message-passing Communications on
  Parallel Systems}.
\newblock {\em Journal of Supercomputing}, 60(1):117--140, 2012.

\bibitem{wadler92}
Philip Wadler.
\newblock The essence of functional programming.
\newblock In {\em Principles of Programming Languages}, pages 1--14. ACM, 1992.

\bibitem{wadler98}
Philip Wadler.
\newblock Why no one uses functional languages.
\newblock {\em SIGPLAN Not.}, 33(8):23--27, August 1998.

\bibitem{spark10}
Matei Zaharia, Mosharaf Chowdhury, Michael~J. Franklin, Scott Shenker, and Ion
  Stoica.
\newblock Spark: cluster computing with working sets.
\newblock In {\em USENIX conference on Hot topics in cloud computing}, page~10.
  USENIX Association, 2010.

\end{thebibliography}

\end{document}